\documentclass[a4paper,11pt]{article}
\pdfoutput=1

\usepackage{jheppub}
\usepackage[T1]{fontenc}

\usepackage{amssymb, amsmath, amsthm}
\usepackage{graphicx}
\usepackage{enumitem}
\usepackage{tikz}
\usetikzlibrary{calc}

\usepackage{longtable, multirow, booktabs}
\usepackage[thicklines]{cancel}

\newenvironment{bsmallmatrix}{
    \left[\begin{smallmatrix}
}{
    \end{smallmatrix}\right]
}

\DeclareMathOperator{\GCD}{gcd}

\DeclareMathOperator{\ns}{null}
\DeclareMathOperator{\rank}{rank}

\usepackage{ifthen}
\newcommand{\bi}[3]{
     \ifthenelse{\equal{#1}{0} \AND \equal{#2}{0} \AND \equal{#3}{0}}{$(0,0,0)$}
    {\ifthenelse{\equal{#1}{0} \AND \equal{#2}{0} \AND \equal{#3}{1}}{$(0,0,\tfrac{1}{2})$}
    {\ifthenelse{\equal{#1}{0} \AND \equal{#2}{1} \AND \equal{#3}{0}}{$(0,\tfrac{1}{2},0)$}
    {\ifthenelse{\equal{#1}{1} \AND \equal{#2}{0} \AND \equal{#3}{0}}{$(\tfrac{1}{2},0,0)$}
    {\ifthenelse{\equal{#1}{0} \AND \equal{#2}{1} \AND \equal{#3}{1}}{$(0,\tfrac{1}{2},\tfrac{1}{2})$}
    {\ifthenelse{\equal{#1}{1} \AND \equal{#2}{0} \AND \equal{#3}{1}}{$(\tfrac{1}{2},0,\tfrac{1}{2})$}
    {\ifthenelse{\equal{#1}{1} \AND \equal{#2}{1} \AND \equal{#3}{0}}{$(\tfrac{1}{2},\tfrac{1}{2},0)$}
    {\ifthenelse{\equal{#1}{1} \AND \equal{#2}{1} \AND \equal{#3}{1}}{$(\tfrac{1}{2},\tfrac{1}{2},\tfrac{1}{2})$}
    {}}}}}}}}
}

\title{134 Billion Intersecting Brane Models}

\author{Gregory J.\ Loges}
\author{and Gary Shiu}

\affiliation{
    Department of Physics, University of Wisconsin-Madison,\\
    1150 University Ave, Madison WI 53706, USA
}

\emailAdd{gloges@wisc.edu}
\emailAdd{shiu@physics.wisc.edu}

\abstract{
The landscape of string vacua is very large, but generally expected to be finite in size.
Enumerating the number and properties of the vacua is an important task for both the landscape and the swampland, in part to gain a deeper understanding of what is possible and ``generic''.
We obtain an {\it exact} counting of distinct intersecting brane vacua of type IIA string theory on the $\mathbb{T}^6/\mathbb{Z}_2\times\mathbb{Z}_2$ orientifold. Care is taken to only count gauge-inequivalent brane configurations.
Leveraging the recursive nature by which branes may be added together one-by-one, we use dynamic programming to efficiently count the number of solutions of the tadpole, K-theory and supersymmetry consistency conditions. The distributions of 4D gauge group rank and complex structure moduli for the entire ensemble of intersecting brane vacua are presented.
The methods we developed here may be useful in obtaining sharp upper and lower bounds on other corners of the landscape.
}

\begin{document}

\maketitle


\section{Introduction}
\label{sec:intro}

The string landscape is enormous. The large number of string vacua comes from myriad choices for the internal geometry, brane configurations and quantized fluxes.
Nevertheless, the expectation is that the landscape of string theory (or more generally quantum gravity) is finite \cite{Vafa:2005ui, Acharya:2006zw,Douglas:2010ic}, in the sense that there are only finitely many\footnote{after quotienting by moduli spaces} low energy effective field theories with a fixed, finite energy cutoff that are consistent with quantum gravity.
This notion of finiteness was argued for using the finiteness of black hole entropy in~\cite{Hamada:2021yxy} and rephrased in terms of tamed geometry in~\cite{Grimm:2021vpn}.

Efforts to estimate the size of the landscape can be roughly split into two categories. On one hand, one can try to understand the number and properties of background geometries, e.g.\ classes of Calabi-Yau 3-folds. On the other hand, for a fixed geometry one can investigate the choices for fluxes and branes which give consistent, controllable vacua. While there has been much work in finding upper bounds on the number of vacua (e.g.\ see~\cite{Ashok:2003gk,Taylor:2015xtz,Demirtas:2020dbm}), it is equally important to have (non-trivial) lower bounds in order to have a true grasp of the landscape's actual extent. In this work we do this precisely for type IIA string theory on the $\mathbb{T}^6/\mathbb{Z}_2\times\mathbb{Z}_2$ orientifold by providing the \emph{exact} number of vacua. The employed algorithms may be adapted and used to obtain exact counts of the number of vacua in other string theoretical settings or at the very least provide a way to meaningfully bound the size of the landscape from below.

The choice of background geometry for our investigation was motivated by results of a number of previous works. Initial explorations of the $\mathbb{T}^6/\mathbb{Z}_2\times\mathbb{Z}_2$ orientifold in~\cite{Cvetic:2001nr,Cvetic:2001tj,Cvetic:2002pj} unearthed several MSSM-like brane configurations (see also~\cite{Berkooz:1996km,Aldazabal:2000dg,Aldazabal:2000cn,Ibanez:2001nd,Blumenhagen:2001te,Blumenhagen:2005mu,ibanez2012}). The number of solutions to the self-consistency equations for this compactification geometry was shown to be finite in~\cite{Douglas:2006xy}, although a sharp bound was not achieved. Landscape statistics were studied in~\cite{Gmeiner:2005vz,Blumenhagen:2004xx} for relatively small values of the moduli and winding numbers. As we shall see, with our way of performing an exact count,
we can bypass the difficulty of scanning over a range of moduli values and/or winding numbers.
These earlier studies of landscape statistics have also estimated the frequency of an MSSM like gauge group with three generations to be one in a billion. A result of the present work is to establish that the number of distinct solutions is at least in the billions. Recently, more targeted brute-force searches have been completed for Pati-Salam realizations~\cite{He:2021gug,He:2021kbj,Li:2022fzt}, where an exhaustive list of $\mathcal{O}(30)$ phenomenologically interesting solutions are found. Machine learning has also come to bear in generating solutions and searching for interesting configurations~\cite{Halverson:2019tkf,Loges:2021hvn} (see also~\cite{Carifio:2017bov,He:2017aed,Ruehle:2017mzq,Mutter:2018sra,Cole:2019enn,Deen:2020dlf,Otsuka:2020nsk,Krippendorf:2021uxu,Berman:2021mcw,Berglund:2021ztg} for applications of machine learning to other areas of the string landscape). In this paper we complete the earlier attempt of~\cite{Douglas:2006xy} by finding the exact number of inequivalent vacua.

\medskip

The rest of this paper is organized as follows. In Sec.~\ref{sec:orbifoldreview} we review the model at hand and its consistency conditions. In Sec.~\ref{sec:uniquevacua} we discuss the symmetries which relate nominally different solutions and how double-counting of vacua is avoided. In Sec.~\ref{sec:counting} we outline the process by which an exhaustive list of moduli is produced and describe how dynamic programming may be used to efficiently count the number of vacua corresponding to each set of moduli. Finally, we conclude in Sec.~\ref{sec:disc}. Details of the algorithms used are relegated to the Appendices.


\section{\texorpdfstring{The $\mathbb{T}^6/\mathbb{Z}_2\times\mathbb{Z}_2$ orientifold}{The T6/Z2xZ2 orientifold}}
\label{sec:orbifoldreview}

In this section we briefly review type IIA string theory on the toroidal orbifold $\mathbb{T}^6/\mathbb{Z}_2\times\mathbb{Z}_2$ with orientifold projection (see, e.g.,~\cite{Lust:2004ks, Blumenhagen:2005mu, Blumenhagen:2006ci, ibanez2012} for a more detailed discussion). Writing $z^i$ ($i=1,2,3$) for complex coordinates on $\mathbb{T}^6$, the $\mathbb{Z}_2\times\mathbb{Z}_2$ action is generated by
\begin{equation}\label{eq:orbifold}
\begin{aligned}
	\theta:\quad (z^1,z^2,z^3) &\mapsto (z^1,-z^2,-z^3) \,,\\
	\omega:\quad (z^1,z^2,z^3) &\mapsto (-z^1,z^2,-z^3) \,.
\end{aligned}
\end{equation}
Quotienting by this group action reduces the number of moduli; in particular, the 6-torus factorizes as $\mathbb{T}^6=\mathbb{T}^2\times\mathbb{T}^2\times\mathbb{T}^2$. The orientifold action is $(-1)^F\Omega\overline{\sigma}$, where $F$ and $\Omega$ are world-sheet fermion number and parity, respectively, and $\overline{\sigma}$ is an anti-holomorphic involution acting as complex conjugation, $\overline{\sigma}(z^i)\to(z^i)^\ast$. For each 2-torus we can write the identifications as
\begin{equation}
	z^i \sim z^i + (1 + ib_i) \sim z^i + i \,,
\end{equation}
where compatibility with the orientifold projection requires $b_i\in\{0,\frac{1}{2}\}$. O6-planes reside at the fixed points of $\overline{\sigma}$, $\overline{\sigma}\theta$, $\overline{\sigma}\omega$ and $\overline{\sigma}\theta\omega$ and introduce negative Ramond--Ramond charge that must be canceled by the inclusion of D6-branes along with their orientifold images. This amounts to the homological condition
\begin{equation}
	\sum_aN_a\big([\Pi_a] + [\Pi_{a'}]\big) - [\Pi_\text{O6}] = 0 \,,
\end{equation}
where $[\Pi_a]$ denotes the homology class of the 3-cycle $\Pi_a$ wrapped by the $N_a$ D6-branes in stack $a$ (the orientifold image is denoted $a'$) and $[\Pi_\text{O6}]$ denotes the total homology class of the O6-planes. We restrict attention to 3-cycles which factorize into three 1-cycles, one wrapping each 2-torus factor, which are labeled by pairs of co-prime winding numbers, $\otimes_{i=1}^3(n_a^i,m_a^i)$. Each brane stack $a$ is thus described by a stack size $N_a$ and six winding numbers which are conveniently collected into the combinations
\begin{equation}
\begin{aligned}
	\widehat{X}_a^0 &= n_a^1n_a^2n_a^3 \,, & \widehat{X}_a^1 &= -n_a^1\widehat{m}_a^2\widehat{m}_a^3 \,,\quad & \widehat{X}_a^2 &= -\widehat{m}_a^1n_a^2\widehat{m}_a^3 \,,\quad & \widehat{X}_a^3 &= -\widehat{m}_a^1\widehat{m}_a^2n_a^3 \,,\\
	\widehat{Y}_a^0 &= \widehat{m}_a^1\widehat{m}_a^2\widehat{m}_a^3 \,, \quad & \widehat{Y}_a^1 &= -\widehat{m}_a^1n_a^2n_a^3 \,, & \widehat{Y}_a^2 &= - n_a^1\widehat{m}_a^2n_a^3 \,, & \widehat{Y}_a^3 &= -n_a^1n_a^2\widehat{m}_a^3 \,,
\end{aligned}
\end{equation}
where $\widehat{m}_a^i = m_a^i + 2b_i(n_a^i+m_a^i)\in\mathbb{Z}$ and $\GCD(n_a^i,m_a^i)=1$.\footnote{Note that in our conventions $\widehat{m}^i$, $\widehat{X}^I$ and $\widehat{Y}^I$ are always integers.} In terms of these the consistency conditions required of the vacua are
\begin{equation}
	\begin{aligned}
		\text{Tadpole:}& & \quad\sum_aN_a\widehat{X}_a^I &= T \,, & \qquad\qquad\text{SUSY--X:}& & \quad\sum_{I=0}^3\widehat{X}_a^I\widehat{U}_I &> 0 \,,\\
		\text{K-theory:}& & \sum_aN_a\widehat{Y}_a^I &\in 2\mathbb{Z} \,, & \text{SUSY--Y:}& & \sum_{I=0}^3\frac{\widehat{Y}_a^I}{\widehat{U}_I} &= 0 \,.
	\end{aligned}
\end{equation}
Given the number of O6-planes, the physically relevant choice is $T=8$ but we keep $T$ arbitrary at some points going forward. The moduli are $\widehat{U}_0=R_x^1R_x^2R_x^3$, $\widehat{U}_1=(1-b_2)(1-b_3)R_x^1R_y^2R_y^3$, etc., where $R_{x,y}^i$ are the 2-tori radii, and must be strictly positive. The volume modulus is given by $\mathcal{V}^2\sim\widehat{U}_0\widehat{U}_1\widehat{U}_2\widehat{U}_3$ and the three complex structure moduli are $\widehat{U}_i/\widehat{U}_0$.

Each vacuum state amounts to specifying a choice for $b_i$ and $k$ sets of winding numbers and stack sizes for which all $2k+8$ consistency conditions above are satisfied for some values of $\widehat{U}_I>0$. Some phenomenologically relevant properties can be determined from the topological data alone, including the gauge group and chiral spectrum. However, with the techniques we use in Sec.~\ref{sec:counting} to count vacua we have limited access to this phenomenological data. As we will describe below, we can track the rank of the gauge group without too much trouble.

Brane configurations satisfying the SUSY--X and SUSY--Y conditions are necessarily composed only of so-called type $A$, $B$ and $C$ branes which have 4, 2 and 1 nonzero tadpoles $\widehat{X}_a^I$, respectively. To satisfy both SUSY conditions, type $A$ branes must have exactly one negative tadpole and we will use $A_J$ to refer to $A$ branes for which $\widehat{X}_a^J$ is negative. Similarly, $B_{JK}$ branes have $\widehat{X}_a^J,\widehat{X}_a^K>0$ and $C_J$ branes have $\widehat{X}_a^J>0$. In our normalization $C$ branes each contribute $2^\nu$ to their respective tadpoles, where $\nu\equiv\sum_i2b_i$ is the number of tilted tori, and importantly have $\widehat{Y}_a^I=0$ so that neither the K-theory nor SUSY conditions are altered by their inclusion.


\section{Vacua, moduli \& counting schemes}
\label{sec:uniquevacua}

What should qualify as distinct vacua? There are three effects that need to be addressed: (i) nominally different sets of winding numbers and tori shapes can describe the same physical brane configurations, (ii) for a given brane configuration there can be different choices for the moduli $\widehat{U}_I$, and (iii) one has to choose how to count configurations which have multiple branes wrapping the same 3-cycle. Let us discuss each of these points in turn.

\paragraph{i. Residual symmetries}
The encoding of a brane configuration into a collection of winding numbers, stack sizes and torus tilts is not unique. There remains the group of large diffeomorphism of $\mathbb{T}^6/\mathbb{Z}_2\times\mathbb{Z}_2\times\mathbb{Z}_2^\mathcal{O}$ which relate nominally different vacua, the most obvious of which is the freedom to relabel the three 2-tori, i.e.\ permute the values of $i$. These large diffeomorphism fall into categories associated with the orientifold action, the orbifold action and the $\mathbb{T}^2\times\mathbb{T}^2\times\mathbb{T}^2$ covering space:
\begin{itemize}
    \item Orientifold-equivalent windings,
    \begin{equation}
        \rho_\sigma:\;(n^i,\widehat{m}^i)\mapsto(n^i,-\widehat{m}^i) \,,
    \end{equation}
    under which $(\widehat{X}^I,\widehat{Y}^I)\mapsto(\widehat{X}^I,-\widehat{Y}^I)$ (branes and their orientifold images are interchanged).
    \item Orbifold-equivalent windings (cf.~Eqn.~\eqref{eq:orbifold}),
    \begin{equation}
    \begin{aligned}
        \rho_\theta:\; (n^i,\widehat{m}^i) &\mapsto (n^1,\widehat{m}^1)\otimes(-n^2,-\widehat{m}^2)\otimes(-n^3,-\widehat{m}^3) \,,\\
        \rho_\omega:\; (n^i,\widehat{m}^i) &\mapsto (-n^1,-\widehat{m}^1)\otimes(n^2,\widehat{m}^2)\otimes(-n^3,-\widehat{m}^3) \,,
    \end{aligned}
    \end{equation}
    under which $\widehat{X}^I$ and $\widehat{Y}^I$ are unchanged.
    \item Orientation/labelling of $\mathbb{T}^2\times\mathbb{T}^2\times\mathbb{T}^2$,
    \begin{equation}
    \begin{aligned}
        \rho_{(01)}:\; (n^i,\widehat{m}^i) &\mapsto (-n^1,-\widehat{m}^1)\otimes(\widehat{m}^3,-n^3)\otimes(\widehat{m}^2,-n^2) & \text{and}\;\; (b_1,b_2,b_3) &\mapsto (b_1,b_3,b_2) \,,\\
        \rho_{(123)}:\; (n^i,\widehat{m}^i) &\mapsto (n^3,\widehat{m}^3)\otimes(n^1,\widehat{m}^1)\otimes(n^2,\widehat{m}^2) & \text{and}\;\;(b_1,b_2,b_3) &\mapsto (b_3,b_1,b_2) \,,\\
    \end{aligned}
    \end{equation}
    which generate all possible permutations of the index $I$: $\mathfrak{S}_4\cong\langle(01),(123)\rangle$.
\end{itemize}
Under these transformations the consistency conditions for the vacua are preserved. Notice, importantly, that while the above are naturally presented in terms of their action on $\widehat{m}^i$, the implied transformations for $m^i$ always both result in integer winding numbers and preserve the co-primality conditions. For example,
\begin{equation}
    \rho_{(01)}:\; (n^i,m^i) \mapsto (-n^1,-m^1)\otimes(\widehat{m}^3,-n^3-2b_3m^3)\otimes(\widehat{m}^2,-n^2-2b_2m^2)
\end{equation}
gives integer $m^i$ and it is straightforward to show that $\GCD(\widehat{m}^i,n^i+2b_im^i) = \GCD(m^i, n^i)$.

Brane configurations represented in terms of winding numbers, etc., which are gauge-equivalent are mapped to one another under repeated applications of the above transformations. To count only gauge-inequivalent vacua we should count the number of these gauge orbits, or, equivalently, pick a representative from each to be counted. This can be done by (i) using $\rho_\sigma,\rho_\theta,\rho_\omega,\rho_{\theta\omega}$ on each set of winding numbers individually to completely fix their signs, and (ii) using the maps which permute the index $I$ to ensure that $0<\widehat{U}_0\leq\widehat{U}_1\leq\widehat{U}_2\leq\widehat{U}_3$. Of course, after having fixed the moduli in this way we will have to consider all eight choices for $b_i$. A configuration with $n_A$, $n_B$ and $n_C$ branes of type $A$, $B$ and $C$, respectively, belongs to a gauge orbit of size $\mathcal{O}(8^{n_A}8^{n_B}4^{n_C}4!)$. Clearly the number of vacua will be greatly inflated if one does not remove gauge redundancies.

\paragraph{ii. Moduli}
The SUSY conditions necessarily allow multiple choices for the moduli $\widehat{U}_I$ since we can always perform an overall rescaling $\widehat{U}_I\to\lambda\widehat{U}_I$ with $\lambda>0$. There also exist brane configurations for which there are two- or three-parameter families of admissible moduli. This means that attempting to count the number of vacua by scanning over various $\widehat{U}_I$ will double-count solutions. We now describe how to make the correspondence between brane configurations and moduli more precise and in particular avoid all double-counting.

The key observation is that the SUSY--Y constraint can be viewed as a matrix equation:
\begin{equation}
	\sum_{I=0}^3\frac{\widehat{Y}_a^I}{\widehat{U}_I} = 0 \quad\forall a \qquad\implies\qquad \mathbb{M}\widehat{u} = 0 \,, \qquad \widehat{u}_I \equiv 1/\widehat{U}_I \,.
\end{equation}
The matrix $\mathbb{M}$, which we refer to as a ``constraint matrix'', imposes linear constraints on $\widehat{u}$ and the set of admissible moduli are identified with $\ns^+\!(\mathbb{M})$, the intersection of the null space of $\mathbb{M}$ with the positive orthant (since the moduli must be positive). Without loss of generality we can take the matrix $\mathbb{M}$ to be in integer reduced row-echelon form (IRREF)\footnote{An integer matrix is said to be in integer reduced row-echelon form if (i) all non-zero rows occur before rows of all zeros, (ii) each non-zero row's pivot is positive, appears to the right of the previous row's pivot and is the only nonzero number in its column, and (iii) the gcd of each non-zero row is one.} since this form is unique and performing Gauss--Jordan elimination does not change the null space. With $k$ D6-brane stacks $\mathbb{M}$ is a $k\times4$ matrix; however, it is clear that $\mathbb{M}$ may never include more than three linearly independent constraints (else the only solution is $\widehat{u}_I=1/\widehat{U}_I=0$), so we can take $\mathbb{M}$ to always be a $3\times4$ matrix, padded with zeros if necessary.

Let us exemplify this correspondence with a two-stack configuration: suppose there is an $A$ brane with $\widehat{Y}_a^I=(-5,3,3,5)$ and a $B$ brane with $\widehat{Y}_b^I=(3,-4,0,0)$. Then
\begin{equation}
	\mathbb{M} = \operatorname{\mathcal{R}}\left(\begin{bsmallmatrix}
		-5 & 3 & 3 & 5\\
		3 & -4 & 0 & 0\\
		0 & 0 & 0 & 0
	\end{bsmallmatrix}\right) = \begin{bsmallmatrix}
		11 & 0 & -12 & -20\\
		0 & 11 & -9 & -15\\
		0 & 0 & 0 & 0
	\end{bsmallmatrix} \,,
\end{equation}
where $\mathcal{R}$ indicates the process of performing Gauss--Jordan elimination to bring the matrix to IRREF. There is a two-parameter family of allowed moduli:
\begin{equation}
	\widehat{u}_I = (\widehat{U}_I)^{-1} = \lambda\big(12+20\mu,9+15\mu,11,11\mu\big) \,, \qquad \lambda>0 \,, \;\; \mu\in\big[\tfrac{2}{15},1\big] \,.
\end{equation}
Importantly, because the correspondence is between vacua and the \emph{linear subspace} of allowed $\widehat{u}_I$, these two-stack vacua are not associated with \emph{any} rank-3 constraint matrices where the moduli are completely fixed (up to the overall rescaling).

Rather than trying to count solutions corresponding to a fixed set of moduli we will count solutions corresponding to a fixed matrix $\mathbb{M}$, which ensures that vacua are counted only once. Clearly there remains the question of which $3\times4$ matrices should be considered. Since the number of vacua is finite so too is the number of $\mathbb{M}$ to be considered. In discussing the large diffeomorphisms above we showed that one can always take the moduli to be ordered, so we should restrict attention to those $\mathbb{M}$ for which $\ns^+(\mathbb{M})$ intersects the cone $\widehat{u}_0\geq\widehat{u}_1\geq\widehat{u}_2\geq\widehat{u}_3>0$. We refer to a matrix $\mathbb{M}$ as \emph{admissible} if its null space intersects this cone.\footnote{Determining if a matrix $\mathbb{M}$ is admissible can be phrased as a problem in linear optimization: maximize $\widehat{u}_3=1/\widehat{U}_3$ subject to the constraints $\widehat{u}_3\leq\widehat{u}_2\leq\widehat{u}_1\leq\widehat{u}_0=1$ and $\mathbb{M}\widehat{u}=0$. If a solution exists and $\widehat{u}_3^\text{max}>0$, then $\mathbb{M}$ is admissible.} The condition that $\mathbb{M}$ be admissible is enough to completely remove all physically redundant constraint matrices which have rank three, but falls short for rank-1 and rank-2 matrices. If an admissible matrix remains admissible upon a permutation of its columns (corresponding to a permutation of the index $I$), then the two resulting matrices will correspond to physically indistinguishable vacua and one should be removed. We do this in the following systematic way: a rank-1 or rank-2 matrix $\mathbb{M}$ is redundant and thus discarded if there exists a permutation, $\sigma\in\mathfrak{S}_4$, of its columns such that $\mathbb{M}_\sigma\equiv\mathcal{R}(\sigma(\mathbb{M}))$ (i) is also admissible, and (ii) $\mathbb{M}_\sigma < \mathbb{M}$ according to the dictionary order on its entries. In Sec.~\ref{sec:allmatrices} we will outline the method by which an exhaustive list of admissible constraint matrices is constructed.

\paragraph{iii. Counting schemes}
When multiple D6-branes wrap the same 3-cycle ($N_a>1$) there are two natural ways to associate a ``count'' to this stack. We refer to these different choices as ``counting schemes'', denoted $c(N_a)$. The first counting scheme is to say that there is \emph{one} way to have these branes: $c(N_a)=1$. Different positions for the branes are connected by dynamical processes wherein the position moduli move along flat directions. The second counting scheme is to say that there are $c(N_a)=p(N_a)$ ways to have these branes, corresponding to counting the number of ways that the branes can be organized into physically separated substacks ($p(n)$ is the number of partitions of $n$). One could argue that these should be counted separately because the resulting gauge groups are different, exhibiting various symmetry-breaking patterns. This second scheme has some undesirable features inherited from the fact that partitions numbers grow very quickly\footnote{The asymptotic expression $p(n)\sim \frac{1}{4n\sqrt{3}}\exp{\left(\pi\sqrt{\frac{2n}{3}}\right)}$ is due to Hardy and Ramanujan.}; for example, a single $A_0$ brane with $\widehat{X}_a^I=(-512,8,8,8)$ and 520 $C_0$ branes would be counted as
\begin{equation}\label{eq:partitions520}
	p(1)p(520) = p(520) = 6,\!889,\!839,\!175,\!409,\!542,\!385,\!648
\end{equation}
distinct configurations. Nevertheless, the choice $c(n)=p(n)$ serves as an ultimate upper bound on the number of vacua by providing the loosest sense in which brane configurations are distinct.


\section{Counting of vacua}
\label{sec:counting}

In the previous section we discussed how all double-counting of gauge-equivalent vacua can be avoided, in part by focusing attention on the constraint matrices which summarize the SUSY--Y consistency conditions. The process of counting vacua is now split into two steps. First, an exhaustive list of constraint matrices must be built. This is helped by the fact that there can be at most three linearly-independent constraints imposed on the moduli. Second, for each constraint matrix there is the combinatorial problem of counting how many configurations of branes satisfy all of the consistency conditions.

\subsection{Constructing all constraint matrices}
\label{sec:allmatrices}

\begin{figure}[t]
    \centering
    \definecolor{myGreen}{RGB}{230,245,230}
    \definecolor{myRed}{RGB}{170,50,50}
    \definecolor{myBlue}{RGB}{50,50,170}
    \tikzstyle{myNode} = [rectangle, minimum height=1.1cm, text width=2.5cm, text centered, draw=black, thick, fill=myGreen]
    \tikzstyle{myArrow} = [very thick, -stealth]
    \begin{tikzpicture}
        \node[myNode] (0) at (0,0) {$\rank{(\mathbb{M})}=0$};
        \node[myNode] (1B) at (-3.75,-2) {$\rank{(\mathbb{M})}=1$ $B$ only};
        \node[myNode] (2B) at (-3.75,-4) {$\rank{(\mathbb{M})}=2$ $B$ only};
        \node[myNode] (3B) at (-3.75,-6) {$\rank{(\mathbb{M})}=3$ $B$ only};
        \node[myNode] (1A) at (3.75,-2) {$\rank{(\mathbb{M})}=1$ $A$ only};
        \node[myNode] (2A) at (3.75,-4) {$\rank{(\mathbb{M})}=2$ $A$ only};
        \node[myNode] (3A) at (3.75,-6) {$\rank{(\mathbb{M})}=3$ $A$ only};
        \node[myNode] (2AB) at (9,-4) {$\rank{(\mathbb{M})}=2$ $A\;\&\;B$};
        \node[myNode] (3AB) at (9,-6) {$\rank{(\mathbb{M})}=3$ $A\;\&\;B$};

        \draw[myArrow, myBlue] (0.south) + (-0.5,0) |- (1B);
        \draw[myArrow, myBlue] (0.south) + (-0.3,0) |- (2B);
        \draw[myArrow, myBlue] (0.south) + (-0.1,0) |- (3B);

        \draw[myArrow, myRed] (0.south) + (0.5,0) |- (1A);
        \draw[myArrow, myRed] (0.south) + (0.3,0) |- (2A);
        \draw[myArrow, myRed] (0.south) + (0.1,0) |- (3A);

        \node[above right, xshift=8, myBlue] at (1B.east) {$1\!\times\!B$};
        \node[above right, xshift=8, myBlue] at (2B.east) {$2\!\times\!B$};
        \node[above right, xshift=8, myBlue] at (3B.east) {$3\!\times\!B$};

        \node[above left, xshift=-8, myRed] at (1A.west) {$1\!\times\!A$};
        \node[above left, xshift=-8, myRed] at (2A.west) {$(2+)\!\times\!A$};
        \node[above left, xshift=-8, myRed] at (3A.west) {$(3+)\!\times\!A$};

        \draw[myArrow, myBlue] (1A.east) + (0,0.1) to[sloped, edge label=$1\!\times\!B$] (2AB.west);
        \draw[myArrow, myBlue] (1A.east) + (0,-0.1) to[sloped, edge label=$2\!\times\!B$, pos=0.6] ($(3AB.west) + (0,0.1)$);
        \draw[myArrow, myBlue] (2A.east) to[sloped, edge label'=$1\!\times\!B$] ($(3AB.west) + (0,-0.1)$);
    \end{tikzpicture}
    \caption{Process by which $A$ and $B$ branes are combined to find an exhaustive list of constraint matrices. Note that the same matrix may be found following different paths. Adding one or two $B$ branes to ``$A$ only'' matrices is easily parallelized.}
    \label{fig:matrixflowchart}
\end{figure}
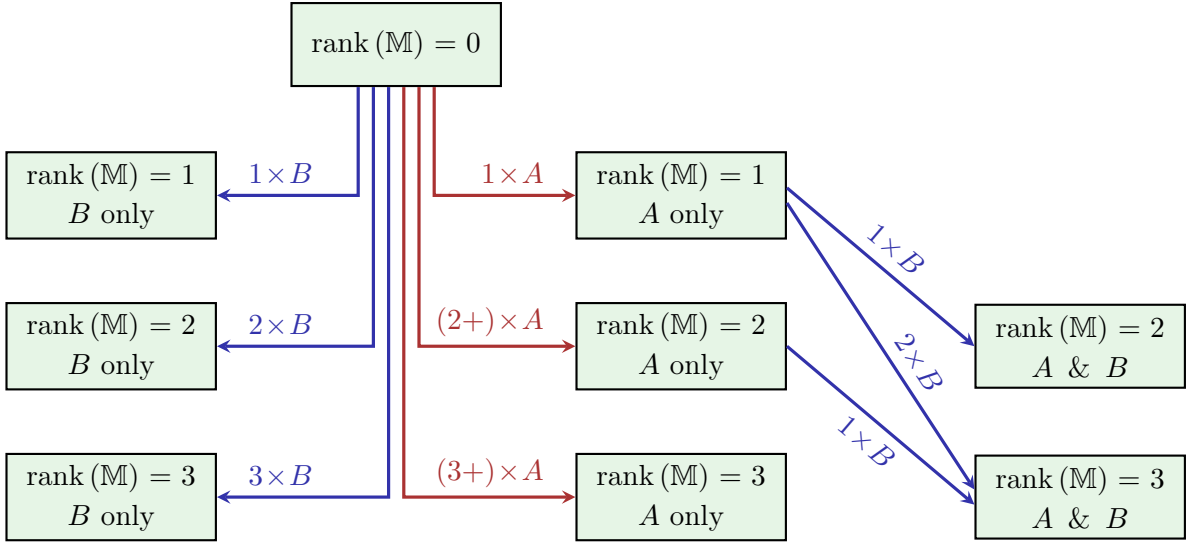

\begin{figure}[t]
    \centering
    \includegraphics[width=\textwidth]{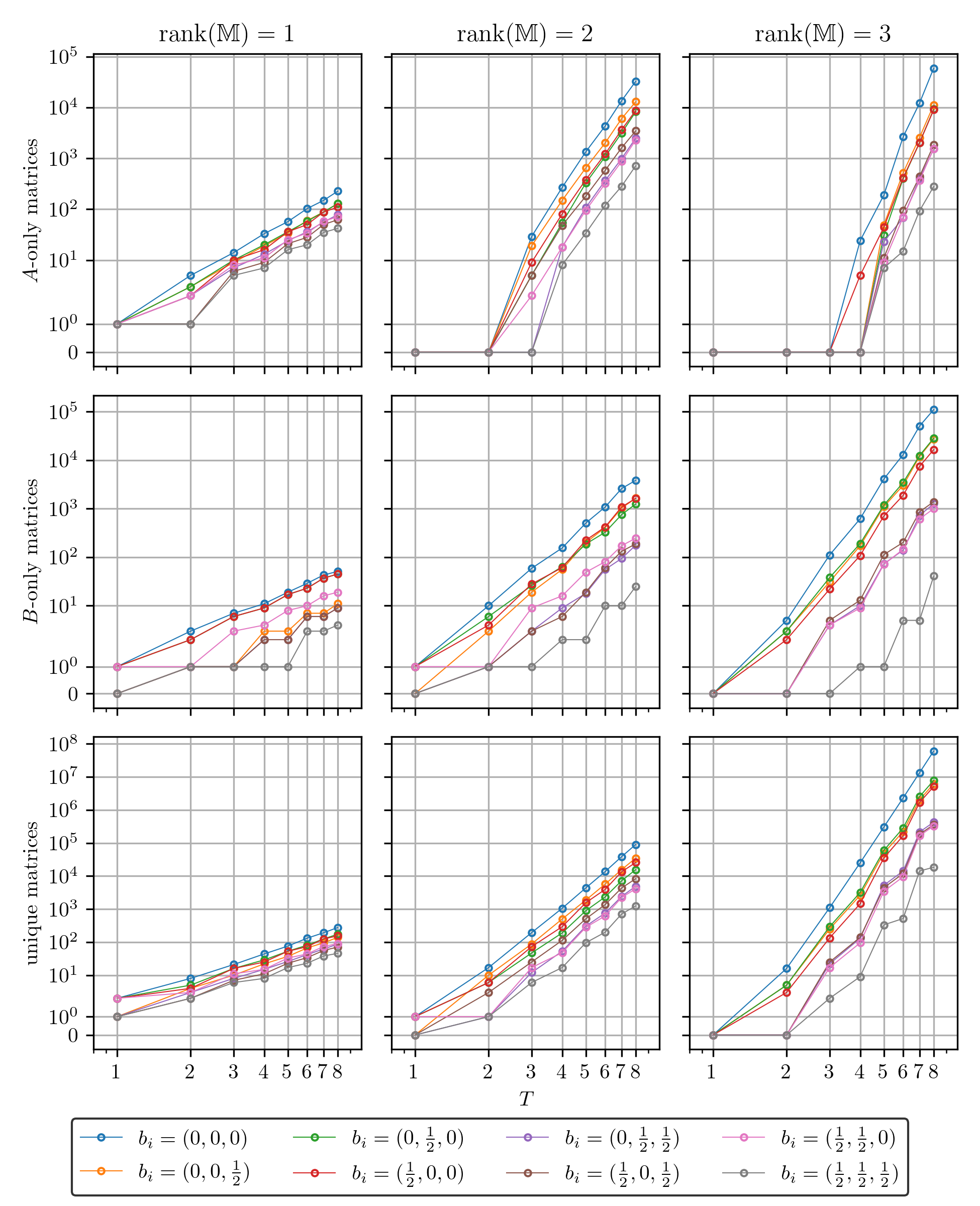}
    \caption{Number of constraint matrices for $T\leq8$ and the eight choices for $b_i$, organized by rank. The first two rows show the number of $A$-only and $B$-only matrices while the third row shows the number of unique matrices (see Fig.~\ref{fig:matrixflowchart}). The numbers of matrices for the same number of tilted tori are nearly coincident. There is also always the single rank-0 matrix $\mathbb{M}=0_{3\times4}$ which corresponds to configurations with only $C$ branes.}
    \label{fig:matrices_grid}
\end{figure}

To create an exhaustive list of constraint matrices our focus is on which combinations of constraints on the moduli can possibly arise from consistent vacua. We look for combinations of only $A$ and $B$ branes which do not over-saturate the tadpoles, ignoring both the K-theory constraints and also that $C$ branes contribute tadpole $2^\nu>1$ when there are tilted tori. More importantly, since at most three linearly-independent constraints can be imposed on the moduli we can limit the search space dramatically to a small number of stacks.

Constraint matrices are found following the strategy outlined in Fig.~\ref{fig:matrixflowchart}, where $A$ and $B$ branes are successively added according to the current value of $\rank{(\mathbb{M})}$. When adding $B$ branes it is straightforward to restrict attention to configurations satisfying the tadpole bounds, but with $A$ branes it is less clear how far to search since some tadpoles are negative. In App.~\ref{app:Atad_bounds} we recall sharp bounds on some of the individual positive and negative tadpoles,
\begin{equation}
    T_-^2 \leq T \,, \qquad T_+^2\leq 2T \,, \qquad T_-^3 \leq \frac{T}{2} \,, \qquad T_+^3 \leq \frac{3T}{2} \,,
\end{equation}
as well as the inequalities
\begin{equation}
    \widehat{X}_a^I > |\widehat{X}_a^J| \,, \qquad \forall\; A_J\text{ branes and }I<J \,,
\end{equation}
which allow one to ``bootstrap'' to more stringent tadpole bounds (see App.~\ref{app:bootstrap}). Using these (bootstrapped) bounds we are able to find all constraint matrices which result from combining only $A$ branes (red arrows in Fig.~\ref{fig:matrixflowchart}) and which do not over-saturate the tadpoles: see App.~\ref{app:algorithms} for details.

After having found all ``$A$ only'' matrices, $B$ branes can be added to each in a straightforward way. Since no more $A$ branes with negative tadpoles are to be added, we need only take the stack sizes for the $B$ branes to be $N_a=1$. In addition, since many $B$ branes can have proportional $\widehat{Y}^I$ we can group the $B$ branes by the constraints they impose on the moduli and treat them collectively. As noted above, since at most three linearly independent constraints can be imposed, the number of $B$ brane constraints to be added is limited.

The number of matrices this process results in is presented in Fig.~\ref{fig:matrices_grid}. In the first two rows are shown the number of matrices at the intermediate steps of Fig.~\ref{fig:matrixflowchart} where only $A$ or only $B$ brane constraints have been incorporated so far. The last row shows the number of unique matrices after collecting all of those produced along all paths of Fig.~\ref{fig:matrixflowchart}. By far the largest number of matrices result from adding two $B$ brane constraints to a single $A_0$ stack; empirically the number of unique, rank-3 matrices grows as $\mathcal{O}(T^{11})$ and dominates over the number of matrices with $\rank{(\mathbb{M})}\leq2$. We emphasize that some matrices will be found multiple times (but repeats are easily removed) and that not all will necessarily correspond to consistent vacua once K-theory and tadpole cancellation are imposed.

\subsection{Counting vacua \& dynamic programming}

For each constraint matrix the number of vacua may be counted independently. With the moduli (partially) fixed it is relatively easy to find a complete list of winding numbers which satisfy the SUSY--Y condition. There then remains the combinatorial task of counting all combinations which satisfy the other consistency conditions.

To accomplish this task it is useful to use dynamical programming, wherein a recursively-defined function or sequence may be computed efficiently by building up values iteratively from the base case(s). As a familiar example, the Fibonacci numbers,
\begin{equation}
    F_n = F_{n-1} + F_{n-2} \,, \qquad F_0=0 \,, \quad F_1=1 \,,
\end{equation}
satisfy a simple recurrence. A na\"ive implementation of this definition would require exponentially many additions to compute $F_n$, ultimately calculating each $F_{k<n}$ many times along the way as the recursion branches out into numerous sub-problems. Of course this is very inefficient and a much better way is to compute and store all of the values of $F_k$ as you go, building up from $k=2$ to $k=n$ one-by-one: this takes only $n-1$ additions.

If we are interested in only \emph{counting} the number of solutions rather than \emph{enumerating} a complete list of configurations and their properties we can leverage dynamic programming to efficiently tally consistent vacua.\footnote{Another simple example of dynamic programming is for the familiar problem of counting the number of permutations of $n$ objects. \emph{Enumerating} all permutations and then counting them quickly becomes unfeasible, but \emph{counting} the number of permutations is quite easy using the recursion $n!=n\times(n-1)!$ with $0!=1$.} The recursion arises from the process by which branes are added to candidate configurations one-by-one, altering the tadpoles, K-theory charges and moduli constraints.

\begin{table}[t]
    \centering
    \begin{tabular}{crrrr}
        & \multicolumn{2}{c}{$\mathbf{c(N_a)=1}$} & \multicolumn{2}{c}{$\mathbf{c(N_a)=p(N_a)}$}\\ \cmidrule(lr){2-3}\cmidrule(lr){4-5}
        $\mathbf{b_i}$ & \multicolumn{1}{c}{\textbf{\cancel{K-theory}}} & \multicolumn{1}{c}{\textbf{K-theory}} & \multicolumn{1}{c}{\textbf{\cancel{K-theory}}} & \multicolumn{1}{c}{\textbf{K-theory}}\\ \cmidrule(r){1-1}\cmidrule(lr){2-3}\cmidrule(l){4-5}

    	\bi{0}{0}{0} &  890,565,159  &  5,554,303  &  20,114,075,740,796,013,040,830  &  134,474,650,261  \\
    	\bi{0}{0}{1} &      478,934  &    478,934  &                      95,430,361  &       95,430,361  \\
    	\bi{0}{1}{0} &      453,539  &    453,539  &                      72,703,621  &       72,703,621  \\
    	\bi{1}{0}{0} &      371,308  &    371,308  &                      62,928,907  &       62,928,907  \\
    	\bi{0}{1}{1} &       19,057  &     19,057  &                         227,872  &          227,872  \\
    	\bi{1}{0}{1} &       16,456  &     16,456  &                         250,768  &          250,768  \\
    	\bi{1}{1}{0} &       14,148  &     14,148  &                         202,210  &          202,210  \\
    	\bi{1}{1}{1} &        1,658  &      1,658  &                           5,899  &            5,899
    \end{tabular}
    \caption{Exact number of vacua with $0<\widehat{U}_0\leq\widehat{U}_1\leq\widehat{U}_2\leq\widehat{U}_3$ for $T=8$, both counting schemes and with K-theory charge cancellation ignored and imposed.}
    \label{tab:T=8counts}
\end{table}

To this end, let $\alpha$ denote a set of branes (i.e.\ a collection of ``gauge-fixed'' winding numbers) and define
\begin{equation}
	F_\alpha[T^I,K^I,\mathbb{M};b_i,c] = \left\{\begin{minipage}{9cm}
		Number of configurations with tilts $b_i$, using the counting scheme $c$ and comprised only of branes from the set $\alpha$ for which $\sum_a N_a\widehat{X}_a^I = T^I$, $\sum_a N_a\widehat{Y}_a^I = K^I$ and the constraints (in IRREF) are $\mathbb{M}$
	\end{minipage}\right\} \,.
\end{equation}
Recall that we are considering both $c=1$ and $c=p$. The K-theory sums and $K^I$ are always to be understood modulo two. As advertised, the function $F$ satisfies the recursion relation
\begin{equation}\label{eq:recursion}
\begin{aligned}
	F_{\alpha\cup \beta}[T^I,K^I,\mathbb{M};b_i,c] &= \big(F_\alpha\ast F_\beta\big)[T^I,K^I,\mathbb{M};b_i,c]\\
	&= \sum_{\substack{T_\alpha^I+T_\beta^I=T^I\\K_\alpha^I+K_\beta^I=K^I\\ \mathcal{R}(\mathbb{M}_\alpha,\mathbb{M}_\beta)=\mathbb{M}}} F_\alpha[T_\alpha^I,K_\alpha^I,\mathbb{M}_\alpha;b_i,c]\times F_\beta[T_\beta^I,K_\beta^I,\mathbb{M}_\beta;b_i,c]
\end{aligned}
\end{equation}
for any two \emph{disjoint} sets of branes $\alpha$ and $\beta$. The convolution amounts to summing over all possible ways to split the tadpoles, K-theory charges and constraints into two. There are two base cases which apply to empty and singleton sets:
\begin{equation}
\begin{aligned}
	F_\emptyset[T^I,K^I,\mathbb{M};b_i,c] &= \begin{cases}
		1 & T^I=K^I=\mathbb{M}=0\\
		0 & \text{otherwise}
	\end{cases}\\
	F_{\{a\}}[T^I,K^I,\mathbb{M};b_i,c] &= \begin{cases}
		c(N) & T^I=N\widehat{X}_a^I \,, \;\; K^I=N\widehat{Y}_a^I \,, \;\; \mathbb{M}=\mathcal{R}(\widehat{Y}_a^I) \,, \;\;N\in\{1,2,3,\ldots\}\\
		1 & T^I=K^I=\mathbb{M}=0\\
		0 & \text{otherwise}
	\end{cases}
\end{aligned}
\end{equation}
The number of fully consistent vacua associated to $\mathbb{M}$ is then
\begin{equation}
    \mathcal{N}(\mathbb{M};b_i,c) = F_\alpha[(T,T,T,T),(0,0,0,0),\mathbb{M};b_i,c] \,.
\end{equation}
Here $\alpha$ encompasses all branes which are compatible with the constraints of $\mathbb{M}$. Similarly, the number of vacua which are consistent other than the K-theory condition is given by
\begin{equation}
    \mathcal{N}_{\cancel{K}}(\mathbb{M};b_i,c) = \sum_{K^I\in\{0,1\}}F_\alpha[(T,T,T,T),K^I,\mathbb{M};b_i,c] \,.
\end{equation}

The results of this count for $T=8$, aggregating the number of vacua associated with each matrix $\mathbb{M}$, are given in Tab.~\ref{tab:T=8counts}. For $b_i=(0,0,0)$ imposing the K-theory conditions greatly reduces the count because large families of configurations with few $A$ and $B$ branes but upwards of 100 $C$ branes are eliminated, while with one or more tilted tori the K-theory conditions are automatic (see App.~\ref{app:allCounts}). The number of vacua for all $T\leq 8$ are tabulated in App.~\ref{app:allCounts}.

\begin{figure}[t]
    \centering
    \includegraphics[width=\textwidth]{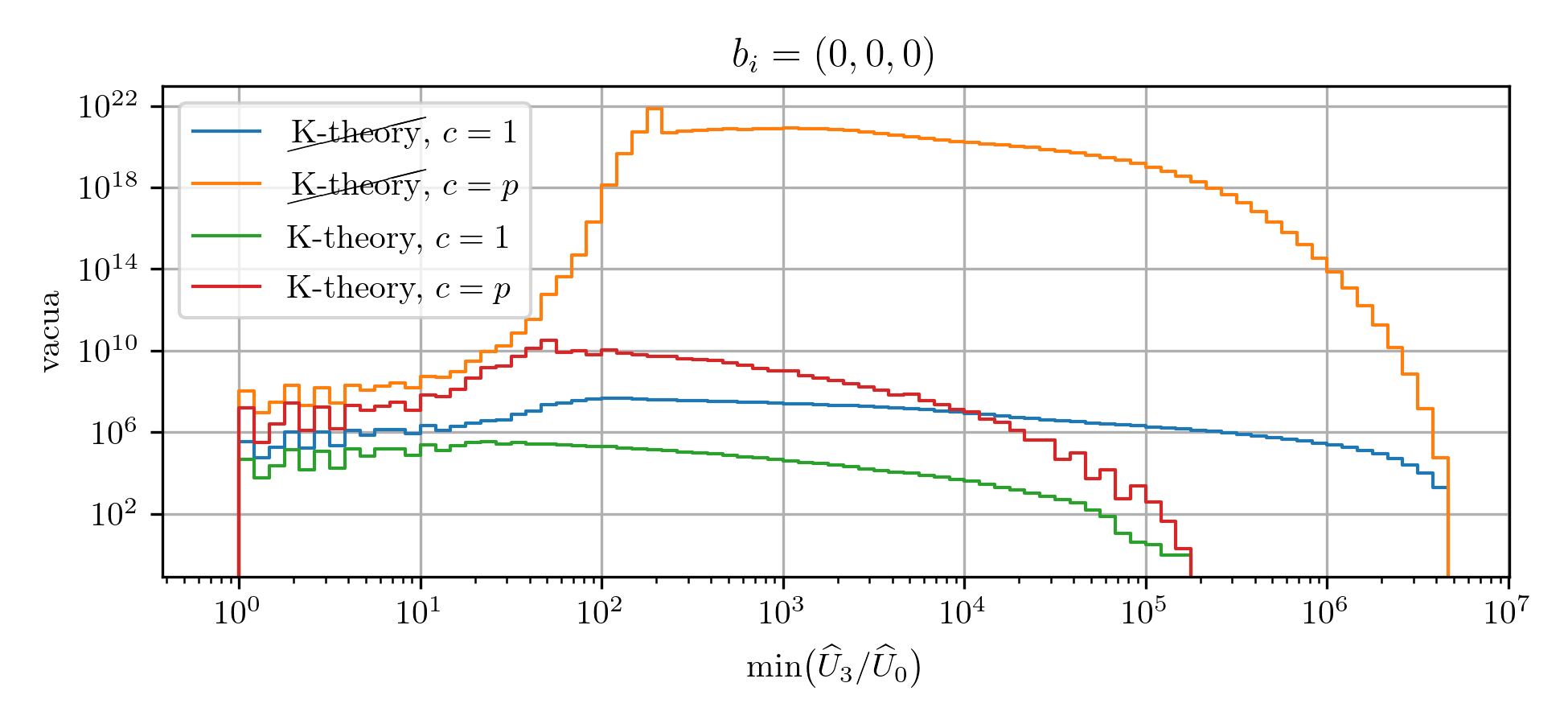}
    \caption{Number of vacua for $T=8$ and $b_i=(0,0,0)$ as a function of the minimal value of $\widehat{U}_3/\widehat{U}_0\geq1$ allowed by the SUSY--Y conditions.}
    \label{fig:rmin_T=8_2bi=000}
\end{figure}

The constraint matrices serve as a proxy for the moduli, and since we can count solutions for each matrix we can also understand how the number of vacua depends on different properties of the $\mathbb{M}$. In Fig.~\ref{fig:rmin_T=8_2bi=000} is given one such distribution, where the number of vacua is shown as a function of $\min(\widehat{U}_3/\widehat{U}_0)$, the smallest value of the largest complex structure modulus allowed by the constraints. For $T=8$ and $b_i=(0,0,0)$ most vacua allow for $\min(\widehat{U}_3/\widehat{U}_0) \approx 10^{1.5\text{--}2.5}$.

\begin{figure}[t]
    \centering
    \includegraphics[width=\textwidth]{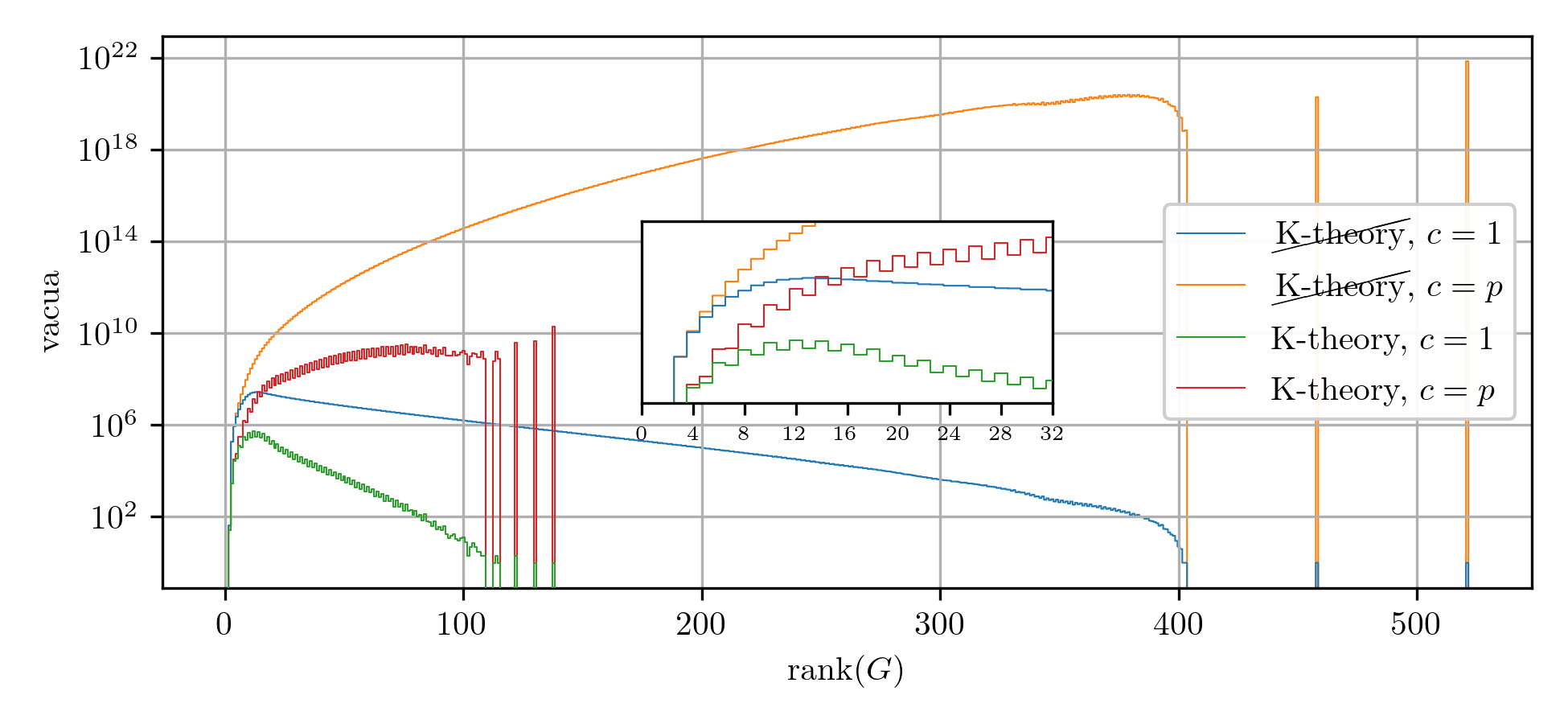}
    \caption{Number of vacua for $T=8$ and $b_i=(0,0,0)$ as a function of the 4D gauge group rank. Note the suppression of odd rank when the K-theory condition is imposed.}
    \label{fig:gaugegroup_T=8_2bi=000}
\end{figure}

While by design the recursion of Eqn.~\eqref{eq:recursion} loses information about which branes combine to give those configurations with particular values for $T^I$, $K^I$ and $\mathbb{M}$, there is some limited phenomenological information that can be extracted. To each $F_\alpha[T^I,K^I,\mathbb{M};b_i,c]$ can be associated a distribution of gauge group ranks which can be updated as branes are added iteratively. This allows one to count the number of vacua of each gauge group rank: see Fig.~\ref{fig:gaugegroup_T=8_2bi=000} for this distribution for $T=8$ and $b_i=(0,0,0)$. The outlier with $\mathrm{rank}(G)=521$ is the example given around Eqn.~\eqref{eq:partitions520}. When K-theory charge cancellation is imposed a large number of vacua are revealed to be inconsistent, including all of those with gauge group rank larger than 138. Also we observe the suppression of odd gauge group rank when the K-theory conditions are imposed. This agrees with the observations of~\cite{Gmeiner:2005vz}, now confirmed to hold on the entire ensemble of consistent vacua.

\section{Discussion}
\label{sec:disc}

In this work we have discussed how the number of intersecting D6-brane vacua for the $\mathbb{T}^6/\mathbb{Z}_2\times\mathbb{Z}_2$ orientifold can be found exactly. There were three main steps to the counting. First, gauge-redundancies were accounted for and explicitly removed by ``fixing a gauge'' for the winding numbers and moduli in Sec.~\ref{sec:uniquevacua}. Next, double-counting of vacua was avoided by shifting attention away from the moduli and onto the \emph{linear subspace} of possible moduli as described by the introduced ``constraint matrices.'' An exhaustive list of these constraint matrices was able to be found without the combinatorics exploding by exploiting that the maximum number of independent constraints is quite small (three) and by leveraging bootstrapped bounds on the positive and negative tadpoles. Finally, because vacua counted for each matrix $\mathbb{M}$ are guaranteed to be distinct, the task of computing the number of fully consistent vacua (using dynamic programming) is easily parallelized.
We note in passing that the main obstacle in getting an exact count of intersecting brane models, in contrast to e.g.\ branes at singularities \cite{Shiu:1998pa,Aldazabal:2000sa,Cvetic:2000st, Berenstein:2001nk, Verlinde:2005jr}, is the presence of negative tadpoles allowed by SUSY branes.

The headline figure is that there are 134 billion gauge-inequivalent vacua for branes wrapping \emph{factorizable} 3-cycles. However, extending the ingredients to include fluxes \cite{Marchesano:2004yq,Marchesano:2004xz} or allowing for brane recombination/non-factorizable cycles would allow for many more vacua and likely a richer phenomenology \cite{Marchesano:2004yn,Feng:2014cla} as well. Taking the one-in-a-billion estimate for MSSM models of~\cite{Gmeiner:2005vz} at face value, we can estimate that there are $\mathcal{O}(100)$ Standard Models in this ensemble. This aligns with the estimate of $\mathcal{O}(10)$ Standard Model vacua in~\cite{Douglas:2006xy} and with the works of~\cite{He:2021gug,He:2021kbj,Li:2022fzt}, where specifying particular features of the gauge group and chiral spectrum results in $\mathcal{O}(30)$ inequivalent vacua. In other settings the number of MSSM realizations is comparable: $\mathcal{O}(100)$ were found for heterotic string theory on the $\mathbb{Z}_6$-II orbifold in~\cite{Lebedev:2006kn} and heterotic CICYs give $\mathcal{O}(200)$ in~\cite{Anderson:2011ns}. In~\cite{Cvetic:2019gnh} an estimate of $\mathcal{O}(10^{15})$ Standard Models coming from F-theory was obtained by estimating the number of triangulations of the bases for a class of elliptically fibered Calabi-Yau fourfolds, although to our knowledge there remains the possibility that gauge-redundancies could reduce this number significantly. It is conceivable that the methods we developed here like dynamic programming can be used to give a meaningful lower bound on the number of distinct Calabi-Yau four-folds.

How does the the total number of vacua -- $\mathcal{O}(10^{11})$ -- compare to other corners of the string theory landscape? A direct comparison is difficult because prior works are mostly limited to estimating upper bounds rather than an exact number. In addition, we have been able to explicitly remove gauge-redundancies. The gauge-orbit sizes determined in Sec.~\ref{sec:uniquevacua}, $\mathcal{O}(8^{n_A}8^{n_B}4^{n_C}4!)$, would easily inflate the number of vacua by tens of orders of magnitude for $n_A+n_B+n_C=\mathcal{O}(T)$. Nevertheless, for comparison, the number of topologically inequivalent Calabi-Yau threefolds coming from the Kreuzer-Skarke list was upper-bounded by $10^{428}$~\cite{Demirtas:2020dbm} and the number of flux vacua estimated to be $10^{272,000}$~\cite{Taylor:2015xtz}. It is an interesting question to develop non-trivial lower bounds or estimate the gauge-orbit volumes for these other counting problems. Using the duality between type IIA orientifolds and $G_2$ compactifications of M theory \cite{Cvetic:2001kk}, we can map the gauge redundancies of brane configurations discussed in this work to that of singularities of $G_2$ manifolds. This suggests that our methods developed here may have their utilities in finding the gauge-orbit volumes for other geometrical constructions of string theory.

The four tadpoles for the toroidal orbifold we study being relatively small allows for a sharp bound on the number of vacua to be found. It would be interesting to extend the used techniques to other backgrounds in order to learn general lessons about how the number of vacua scales with the number and size of the tadpoles. Also, being armed with this knowledge about the entire ensemble of vacua may prove useful in understanding more refined statistics or searching for the needle-in-the-haystack that is the MSSM. See~\cite{Cvetic:2022fnv}
for a recent summary of attempts to construct the MSSM in various corners of the string landscape. For example, machine learning approaches for sampling from this ensemble may be informed by which choices for the moduli allow for more solutions to the consistency conditions. The success of such methods, as measured by the proportion of vacua found, may now be precisely determined. We hope to address these points in the future.


\paragraph{Acknowledgments}
$ $\\

\noindent We thank Andreas Schachner for dicussions. The work of GL and GS is supported in part by the DOE grant DE-SC0017647.
The computations in this research were performed using the compute resources and assistance of the UW-Madison Center For High Throughput Computing (CHTC) in the Department of Computer Sciences. The CHTC is supported by UW-Madison, the Advanced Computing Initiative, the Wisconsin Alumni Research Foundation, the Wisconsin Institutes for Discovery, and the National Science Foundation, and is an active member of the OSG Consortium, which is supported by the National Science Foundation and the U.S.\ Department of Energy's Office of Science.


\appendix

\section{Type A brane tadpole bounds}
\label{app:Atad_bounds}

In this section we review bounds on the tadpole contributions of type $A$ branes, following~\cite{Douglas:2006xy}. While $A$ branes each contribute negatively to one of the four tadpoles and can even have a large enough negative tadpole to make the average tadpole decrease, one can prove that these negative tadpoles cannot grow indefinitely.

First, let us consider an individual $A$ brane, for which
\begin{equation}
    \widehat{Y}_a^I = \frac{1}{\widehat{X}_a^I}\prod_{i=1}^3n_a^i\widehat{m}_a^i \quad\propto\quad \frac{1}{\widehat{X}_a^I} \,.
\end{equation}
The SUSY-Y condition can thus be written as
\begin{equation}
    \sum_I\frac{1}{\widehat{X}_a^I\widehat{U}_I} = 0 \,,
\end{equation}
where exactly one term is negative. For an $A_J$ brane,
\begin{equation}
    \frac{1}{|\widehat{X}_a^J|\widehat{U}_J} = \sum_{I\neq J}\frac{1}{\widehat{X}_a^I\widehat{U}_I} > \frac{1}{\widehat{X}_a^I\widehat{U}_I} \,, \qquad\forall\; I\neq J \,,
\end{equation}
and since (as discussed in Sec.~\ref{sec:uniquevacua}) we can always take $0<\widehat{U}_0\leq\widehat{U}_1\leq\widehat{U}_2\leq\widehat{U}_3$, we conclude that
\begin{equation}\label{eq:Abranetadineq}
    \boxed{\quad\bigg. \widehat{X}_a^I > |\widehat{X}_a^J| \,, \qquad \forall\; A_J\text{ branes and }I<J \,. \quad }
\end{equation}
In particular, we see that only $A_0$ branes can have an \emph{average} tadpole which is non-positive.

Next, let us consider configurations with any number of $A$ branes, for which it is fruitful to split the tadpoles into positive and negative terms and derive bounds on these individually. To this end, define
\begin{equation}
    \widehat{X}_{a,\pm}^I = \frac{|\widehat{X}_a^I| \pm \widehat{X}_a^I}{2} \geq 0 \,, \qquad T_\pm^I = \sum_{a\in A}N_a\widehat{X}_{a,\pm}^I \geq 0 \,.
\end{equation}
The tadpole bounds give
\begin{equation}
    T_+^I - T_-^I = \sum_{a\in A}N_a\big(\widehat{X}_{a,+}^I - \widehat{X}_{a,-}^I\big) = \sum_{a\in A}N_a\widehat{X}_a^I \leq \sum_a N_a\widehat{X}_a^I \leq T \,,
\end{equation}
having used that the tadpole contributions from $B$ and $C$ branes are non-negative. It is also helpful to introduce
\begin{equation}
    \mathcal{G} = \sum_I\sum_{a\in A}N_a\widehat{X}_a^I\widehat{U}_I = \sum_I \big(T_+^I - T_-^I\big)\widehat{U}_I \,,
\end{equation}
which is clearly bounded above as
\begin{equation}
    \mathcal{G} \leq T\sum_I\widehat{U}_I \,.
\end{equation}
A lower bound can be found by swapping the order of the sums, replacing all negative terms using SUSY--X and leveraging the AM--HM inequality\footnote{The AM--HM inequality for three terms is $\frac{x+y+z}{3} \geq \frac{3}{\frac{1}{x}+\frac{1}{y}+\frac{1}{z}}$ for all $x,y,z>0$, with equality only for $x=y=z$.}:
\begin{equation}
\begin{aligned}
    \mathcal{G} &= \sum_{a\in A}N_a\sum_I\widehat{X}_a^I\widehat{U}_I = \sum_J\sum_{a\in A_J}N_a\bigg(\sum_{I\neq J}\widehat{X}_a^I\widehat{U}_I + \widehat{X}_a^J\widehat{U}_J\bigg)\\
    &= \sum_J\sum_{a\in A_J}N_a\bigg(\sum_{I\neq J}\widehat{X}_{a,+}^I\widehat{U}_I - \frac{1}{\sum\limits_{I\neq J} \frac{1}{\widehat{X}_{a,+}^I\widehat{U}_I}} \bigg)\\
    &= \sum_J\sum_{a\in A_J}N_a\sum_{I\neq J}\widehat{X}_{a,+}^I\widehat{U}_I\bigg(1 - \frac{1}{\sum\limits_{I\neq J}\widehat{X}_{a,+}^I\widehat{U}_I\sum\limits_{I\neq J} \frac{1}{\widehat{X}_{a,+}^I\widehat{U}_I}} \bigg)\\
    &\geq \sum_J\sum_{a\in A_J}N_a\sum_{I\neq J}\widehat{X}_{a,+}^I\widehat{U}_I\left(1 - \frac{1}{9} \right) = \frac{8}{9}\sum_J\sum_{a\in A_J}N_a\sum_I\widehat{X}_{a,+}^I\widehat{U}_I = \frac{8}{9}\sum_IT_+^I\widehat{U}_I \,.
\end{aligned}
\end{equation}
That is, we have
\begin{equation}
    \frac{8}{9}\sum_IT_+^I\widehat{U}_I \leq \sum_I\big(T_+^I-T_-^I\big)\widehat{U}_I \leq T\sum_I\widehat{U}_I \,,
\end{equation}
which can be reorganized into the form
\begin{equation}\label{eq:TpTmTinequalities}
    \sum_IT_-^I\widehat{U}_I \leq \frac{1}{9}\sum_IT_+^I\widehat{U}_I \leq \frac{T}{8}\sum_I\widehat{U}_I \,.
\end{equation}
Although this provides an upper bound on the negative tadpoles, the bound is moduli-dependent and there remains the possibility that infinite families of vacua exist with $\widehat{U}_3\gg\widehat{U}_0$. However, using $\widehat{U}_I\leq\widehat{U}_3$ the above inequality \emph{does} provide a moduli-independent bound for $T_\pm^3$~\cite{Douglas:2006xy}:
\begin{equation}\label{eq:T3bounds}
    \boxed{\quad\bigg. T_+^3 \leq \frac{3T}{2} \,, \quad T_-^3 \leq \frac{T}{2} \,. \quad}
\end{equation}
Restricting the sum in $\mathcal{G}$ to $I\leq2$, one can similarly show that~\cite{Douglas:2006xy}
\begin{equation}\label{eq:T2bounds}
    \boxed{\quad\bigg. T_+^2 \leq 2T \,, \quad T_-^2 \leq T \,. \quad}
\end{equation}
Returning to Eqn.~\eqref{eq:TpTmTinequalities}, we can write
\begin{equation}\label{eq:TUbounds}
    \boxed{\quad\Bigg. \sum_I\left(\frac{T}{8} - T_-^I\right)\widehat{U}_I \geq 0 \,, \qquad \sum_I\left(\frac{9T}{8} - T_+^I\right)\widehat{U}_I \geq 0 \,. \quad }
\end{equation}
For $T<8$ we learn that there can be no vacua with all $A_0,A_1,A_2,A_3$ branes coexisting, since then $T_-^I\geq1$ and each term in parentheses in the left above would be negative. It is straightforward to check that such $A$ brane-rich vacua are also forbidden for $T=8$: even though four-stack configurations with $T_-^I=1$ appear to be marginally allowed, there are no combinations that both satisfy the tadpole bound and give an admissible constraint matrix.

Finding moduli-independent bounds for the two remaining negative tadpoles is more involved~\cite{Douglas:2006xy}, but note that just from the above all of the winding numbers $n^i,\widehat{m}^i$ except for $n^1$ are already bounded by some constant times $T$. In practice, configurations with large $T_-^{0,1}\propto n^1$ fall into families which are easy to individually show are finite. For example, for $T=8$ and $b_i=(0,0,0)$ the largest value of $T_-^0$ belongs to the following two-stack family ($N_a=N_v=1$ and $x,y\geq1$):
\begin{align}
    w_a^i &= (x,1)\otimes(3,1)\otimes(-3,-1) \,, \quad& \widehat{X}_a^I &= (-9x,x,3,3) \,, \quad& \widehat{Y}_a^I &= (-1,9,3x,3x)\\
    w_b^i &= (y,-1)\otimes(5,1)\otimes(2,1) \,, & \widehat{X}_b^I &= (10y,-y,5,2) \,, & \widehat{Y}_b^I &= (-1,10,-2y,-5y) \,, \notag
\end{align}
(plus appropriate numbers of $C$ branes). The constraint matrix,
\begin{equation}
    \mathbb{M} = \begin{bsmallmatrix}
        1 & 0 & -30x-18y & -30x-45y\\
        0 & 1 & -3x-2y & -3x-5y\\
        0 & 0 & 0 & 0
    \end{bsmallmatrix} \,,
\end{equation}
is always admissible but the tadpole bounds $10y-9x\leq 8$ and $x-y\leq 8$ only allow out to $(x,y)=(88,80)$, for which $T_-^0=792$. Note, however, that this two-stack family has nonzero K-theory charge since
\begin{equation}
    N_a\widehat{Y}_a^I + N_b\widehat{Y}_b^I \equiv (0, 1, x, x+y) \mod{2} \,,
\end{equation}
and will not survive as fully consistent vacua. Similarly, the largest value of $T_-^1$ appears in the family
\begin{equation}
\begin{aligned}
    N_a &= 1 \,, \qquad& w_a^i &= (x,1)\otimes(1,1)\otimes(-5,-4) \,, \qquad& \widehat{X}_a^I &= (-5x,4x,4,5) \,,\\
    N_b &= 1 \,, & w_b^i &= (y,-1)\otimes(4,3)\otimes(1,1) \,, & \widehat{X}_b^I &= (4y,-3y,4,3) \,.
\end{aligned}
\end{equation}
with the maximal values $(x,y)=(56,72)$, for which $T_-^1=216$.

\subsection{Bootstrapped tadpole bounds}
\label{app:bootstrap}

Stronger bounds on the $A$ brane tadpoles can be found and imposed at intermediate stages of the algorithms to be described in App.~\ref{app:algorithms}. The idea is to bootstrap candidate $A$ brane configurations to larger values of $T_\pm^I$ by repeately using Eqn.~\eqref{eq:Abranetadineq} and then imposing Eqns.~\eqref{eq:T3bounds}, \eqref{eq:T2bounds} and \eqref{eq:TUbounds} to (potentially) rule out the starting configuration. Let us demonstrate how this works in practice with a simple example; suppose we are considering a configuration currently consisting of the $A_1$ and $A_2$ branes with
\begin{equation}
    \widehat{X}_a^I = (3,-2,12,2) \,, \qquad \widehat{X}_b^I = (2,8,-1,4) \,.
\end{equation}
Their constraint matrix,
\begin{equation}
    \mathbb{M} = \begin{bsmallmatrix}
        28 & 0 & -47 & 18\\
        0 & 7 & -9 & -4\\
        0 & 0 & 0 & 0
    \end{bsmallmatrix} \,,
\end{equation}
is admissible and all of the bounds on $T_\pm^{2,3}$ are safely satisfied for $N_a=1$ and $N_b\leq 2$. Nevertheless, these two branes can never coexist in a consistent vacuum, as we now show. Fix $N_b=1$ ($N_b>1$ only exacerbates the issue), so that
\begin{equation}
\begin{aligned}
    T_+^0 &= 5 \,, \qquad&  T_+^1 &= 8 \,, \qquad&  T_+^2 &= 12 \,, \qquad&  T_+^3 &= 6 \,,\\
    T_-^0 &= 0 \,,       &  T_-^1 &= 2 \,,       &  T_-^2 &= 1 \,,        &  T_-^3 &= 0 \,.
\end{aligned}
\end{equation}
Since currently $T_+^2 - T_-^2 = 11 > 8$, a tadpole-cancelling vacuum will be found only if at some point one or more additional $A_2$ branes are included, which according to Eqn.~\eqref{eq:Abranetadineq} must contribute $\Delta T_+^{0,1} > \Delta T_-^2 \geq 3$ and also $\Delta T_+^3\geq 1$. This would bring the tadpoles to
\begin{equation}
\begin{aligned}
    T_+^0 &\geq 9 \,, \qquad&  T_+^1 &\geq 12 \,, \qquad&  T_+^2 &= 12 \,, \qquad&  T_+^3 &\geq 7 \,,\\
    T_-^0 &= 0 \,,          &  T_-^1 &= 2 \,,           &  T_-^2 &\geq 4 \,,     &  T_-^3 &= 0 \,.
\end{aligned}
\end{equation}
At this point we see that both $A_0$ branes (with $\Delta T_-^0\geq1$ and $\Delta T_+^{1,2,3}\geq1$) and additional $A_1$ branes (with $\Delta T_+^0>\Delta T_-^1\geq 2$ and $\Delta T_+^{2,3}\geq1$) would need to be included, bringing the tadpoles to\footnote{Stronger bounds can be found by adding the $A_0$ and $A_1$ tadpole contributions successively.}
\begin{equation}
\begin{aligned}
    T_+^0 &\geq 12 \,, \qquad&  T_+^1 &\geq 13 \,, \qquad&  T_+^2 &\geq 14 \,, \qquad&  T_+^3 &\geq 9 \,,\\
    T_-^0 &\geq 1 \,,        &  T_-^1 &\geq 4 \,,        &  T_-^2 &\geq 4 \,,        &  T_-^3 &= 0 \,.
\end{aligned}
\end{equation}
These values of $T_+^I$ violate Eqn.~\eqref{eq:TUbounds}. Alternatively, since now $T_+^3-T_-^3>8$, adding in the tadpole contributions of $A_3$ brane(s) would give values for $T_-^I\geq 1$ which violate the other half of Eqn.~\eqref{eq:TUbounds}. The initial pair of seemingly promising branes is ruled out by ``bootstrapping'' our way to the larger tadpoles that must exist if they are to be extended to a consistent vacuum. This is all done without reference to the details of the winding numbers, changes to the constraint matrix or particular values of the moduli. Note that, because of the inequalities which appear above, one cannot bootstrap for any value of $I$ more than once.


\section{Algorithm for constructing constraint matrices}
\label{app:algorithms}

In this section we describe the methods used to construct a complete list of constraint matrices. This list will be exhaustive, but may contain matrices which actually do not correspond to \emph{any} vacua since in this section we will ignore the K-theory conditions, $C$ brane tadpoles being larger than one, and $B$ brane multiplicities.

\begin{figure}[t]
	\centering
	\includegraphics[width=\textwidth]{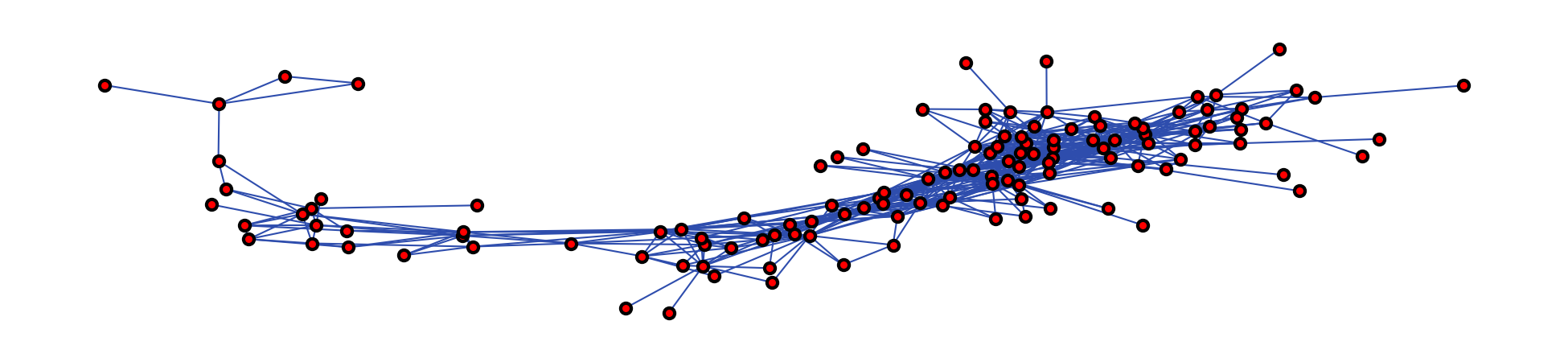}
	\caption{Graph structure for $T=6$ and $b_i=(0,0,0)$, showing the pairwise compatibility of $A_0$ branes with $|\widehat{X}^I|\leq(T^3,T,T,T)$. Each vertex represents an $A_0$ brane and two vertices are connected by an edge if together they give an admissible constraint matrix and satisfy the (bootstrapped) tadpole bounds. There are 399 vertices (270 are isolated and not shown), 539 edges, 364 3-cliques, 107 4-cliques, 20 5-cliques, 1 6-clique and no $k$-cliques for $k\geq7$.}
	\label{fig:graph}
\end{figure}

Each $A$ and $B$ brane corresponds to a constraint imposed on the moduli. When branes are combined these constraints are collected into a constraint matrix $\mathbb{M}$ and brought to IRREF (as described above). Our goal is two-fold: (i) find all combinations of $A$ branes which satisfy the tadpole bounds, and (ii) to these add all possible $B$ branes, all the while restricting attention only to admissible constraint matrices. As shown in App.~\ref{app:Atad_bounds}, all $A_J$ brane types cannot co-exist in a consistent vacuum for $T\leq 8$, and so step (i) breaks into several cases. For each, the negative tadpoles are bounded as $T_-^I<T$ by virtue of Eqn.~\eqref{eq:Abranetadineq} (or $T_-^3\leq\frac{T}{2}$) as long as type $A_0$ branes are not present. The cases and tadpole bounds when $A_0$ branes are present are as follows:
\begin{itemize}
	\item \underline{One or more stacks of $A_J$ branes}\\
	Clearly $T_+^I\leq T$ for all $I$, and $T_-^0 \leq T^3$ since $|\widehat{X}^0| = \left|\frac{\widehat{X}^1\widehat{X}^2\widehat{X}^3}{(\widehat{m}^1\widehat{m}^2\widehat{m}^3)^2}\right| \leq |\widehat{X}^1\widehat{X}^2\widehat{X}^3|$.
	\item \underline{\emph{Exactly} one stack each of $A_J$ and $A_K$ branes, $J<K$}\\
	For $(J,K)=(0,1)$ the largest negative tadpoles grow as $T_-^0\lesssim T^5$, and $T_-^1\lesssim T^3$~\cite{Douglas:2006xy}. For $T=8$ the largest values are 792 and 216, respectively, as discussed in App.~\ref{app:Atad_bounds}. For $(J,K)=(0,2)$ and $(J,K)=(0,3)$ the bound on $T_-^0$ grows as $T_-^0\lesssim T^3$ for the same reason as the previous case. For $K=2$ and $K=3$ the largest values of $T_-^0$ are 48 and 16, respectively.
	\item \underline{One or more stacks each of $A_J$ and $A_K$ branes, $J<K$ (three or more stacks in total)}\\
	The scaling of the bounds on $T_-^0$ and $T_-^1$ are the same as the previous case, but in practice with three or more stacks the extreme values are much lower. For $T=8$ and $(J,K)=(0,1)$ the largest values are $T_-^0=192$ and $T_-^1=64$, and for $(J,K)=(0,2)$ and $(J,K)=(0,3)$ the largest values of $T_-^0$ are 26 and 12, respectively.
	\item \underline{One or more stacks each of $A_J$, $A_K$ and $A_L$ branes, $J<K<L$.}\\
	For $T=8$ and $J=0$ the largest value of $T_-^0$ is 25.
\end{itemize}
The second and third cases are separated only for practical purposes: the negative tadpoles are much smaller when three or more stacks are required and thus the search space is greatly reduced. Let us emphasize that the bootstrapped bounds will be different for each case and choice of $J,K,L$. For example, if looking for configurations with exactly one stack each of $A_1$ and $A_2$ branes, one would bootstrap the tadpoles and impose $T_-^I\leq (0, T-1, T-1, 0)$ and $T_+^I\leq (T, 2T-1,2T-1,T)$.

For each case a complete list of $A_J$ branes satisfying their respective tadpole bounds is quickly generated. A brute-force search of all pairs, triples, quadruples, etc.\ of these $A$ branes, looking for combinations which both give an admissible constraint matrix and satisfy the (bootstrapped) tadpole bounds, is unfeasible. Instead, a nice perspective is given by phrasing the problem at hand in terms of graph theory. Let each $A$ brane correspond to the vertex of a graph and let two vertices be connected by an edge if together these two $A$ branes give an admissible constraint matrix and satisfy the (bootstrapped) bounds on $T_\pm^I$. We are then interested in finding $k$-cliques of this graph, which can be done quickly using the graph's adjacency matrix. For each clique one may then quickly search for combinations which have $T_+^I-T_-^I \leq T$ for all $I$. In the cases where we require certain numbers of stacks of each $A$ brane type, one can refine the above description by coloring vertices based on $J$ and only considering $k$-cliques which contain the required vertex colors. In Fig.~\ref{fig:graph} is shown the graph which results for $T=6$, $b_i=(0,0,0)$ when looking for configurations with one or more stacks of $A_0$ branes (and no others). There are candidate $k$-cliques for $k\leq 6$ which may be quickly scanned through to find the constraint matrices corresponding to configurations with under-saturated tadpoles.

Having found an exhaustive list of matrices which result from combining only $A$ branes, it is then straightforward to add $B$ branes to each in parallel. Since we are only interested in finding the matrices themselves we need only add $B$ branes to matrices which are rank zero, one or two (see Fig.~\ref{fig:matrixflowchart}):
\begin{itemize}
	\item Add up to three $B$ branes to the ``empty'' configuration with no $A$ branes.
	\item Add up to two $B$ branes to rank-1 matrices (which necessarily correspond to only a single $A$ brane stack).
	\item Add exactly one $B$ brane to rank-2 matrices (which correspond to two or more $A$ brane stacks).
\end{itemize}
In addition, $B$ branes can be collected into groups with proportional $\widehat{Y}^I$ which all impose the same constraints on the moduli and treated simultaneously. The resulting list of constraint matrices is exhaustive, meaning that any supersymmetric, tadpole-cancelling vacuum will have its constraint matrix in this list. There will, however, be extraneous matrices which do not correspond to any vacua when tadpole cancellation and the K-theory conditions are imposed (e.g.\ because of stack sizes and $C$ branes having tadpoles larger than one with tilted tori).


\section{Exact numbers of vacua}
\label{app:allCounts}

In Tab.~\ref{tab:allcounts} below we record the exact number of vacua for $T\leq8$ and all eight choices for $b_i$. Results are presented for both counting schemes discussed in Sec.~\ref{sec:uniquevacua} and with the K-theory constraints both ignored and imposed.

A striking feature is that for $b_i\neq(0,0,0)$ the number of solutions either does not change or drops to zero when K-theory is imposed. This is easy to understand as follows. Suppose that $b_i=(0,0,\tfrac{1}{2})$, so that $\widehat{m}^3=2m^3+n^3$. This means that $\widehat{m}^3\equiv n^3\mod{2}$ and also that
\begin{equation}
    \widehat{X}^0+\widehat{Y}^3 = n^1n^2(n^3-\widehat{m}^3) \equiv 0\mod{2}
\end{equation}
Thus we have
\begin{equation}
    K^3 = \sum_aN_a\widehat{Y}_a^3 \equiv \sum_aN_a\widehat{X}_a^0 \mod{2}
\end{equation}
and for a tadpole-cancelling vacuum,
\begin{equation}
    \sum_aN_a\widehat{X}_a^0 = T \qquad\implies\qquad K^3 \equiv T\mod{2}
\end{equation}
(similar calculations show the same holds for \emph{all} $K^I$ whenever there is one or more tilted torus). In particular, if $T$ is odd the K-theory charge can never be cancelled if the tadpole is cancelled and if $T$ is even the K-theory condition follows from tadpole cancellation.

\newpage
\begin{longtable}{ccrrrr}
    \caption{Exact number of vacua with $0<\widehat{U}_0\leq\widehat{U}_1\leq\widehat{U}_2\leq\widehat{U}_3$ for all $T\leq 8$, all choices for $b_i\in\{0,\frac{1}{2}\}$, both counting schemes and with K-theory charge cancellation ignored and imposed.}
    \label{tab:allcounts}\\

    \toprule
    & & \multicolumn{2}{c}{$\mathbf{c(N_a)=1}$} & \multicolumn{2}{c}{$\mathbf{c(N_a)=p(N_a)}$}\\ \cmidrule(lr){3-4}\cmidrule(lr){5-6}
    $\mathbf{T}$ & $\mathbf{b_i}$ & \multicolumn{1}{c}{\textbf{\cancel{K-theory}}} & \multicolumn{1}{c}{\textbf{K-theory}} & \multicolumn{1}{c}{\textbf{\cancel{K-theory}}} & \multicolumn{1}{c}{\textbf{K-theory}}\\ \cmidrule(r){1-2}\cmidrule(lr){3-4}\cmidrule(l){5-6}
    \endfirsthead

    \multicolumn{6}{c}%
    {{\bfseries \tablename\ \thetable{} -- continued from previous page}} \\
    \toprule
    & & \multicolumn{2}{c}{$\mathbf{c(N_a)=1}$} & \multicolumn{2}{c}{$\mathbf{c(N_a)=p(N_a)}$}\\ \cmidrule(lr){3-4}\cmidrule(l){5-6}
    $\mathbf{T}$ & $\mathbf{b_i}$ & \multicolumn{1}{c}{\textbf{\cancel{K-theory}}} & \multicolumn{1}{c}{\textbf{K-theory}} & \multicolumn{1}{c}{\textbf{\cancel{K-theory}}} & \multicolumn{1}{c}{\textbf{K-theory}}\\ \cmidrule(r){1-2}\cmidrule(lr){3-4}\cmidrule(l){5-6}
    \endhead

    \midrule \multicolumn{6}{r}{{Continued on next page}}\\ \bottomrule
    \endfoot

    \bottomrule
    \endlastfoot

	\multirow{8}{*}{1}
	& \bi{0}{0}{0} &  4  &  1  &  5  &  1  \\
	& \bi{0}{0}{1} &  1  &  0  &  1  &  0  \\
	& \bi{0}{1}{0} &  2  &  0  &  2  &  0  \\
	& \bi{1}{0}{0} &  2  &  0  &  2  &  0  \\
	& \bi{0}{1}{1} &  0  &  0  &  0  &  0  \\
	& \bi{1}{0}{1} &  0  &  0  &  0  &  0  \\
	& \bi{1}{1}{0} &  1  &  0  &  1  &  0  \\
	& \bi{1}{1}{1} &  0  &  0  &  0  &  0  \\
	\cmidrule(r){1-2}\cmidrule(lr){3-4}\cmidrule(l){5-6}

	\multirow{8}{*}{2}
	& \bi{0}{0}{0} &  106  &   10  &  237  &   45  \\
	& \bi{0}{0}{1} &    6  &    6  &    9  &    9  \\
	& \bi{0}{1}{0} &    7  &    7  &   14  &   14  \\
	& \bi{1}{0}{0} &    5  &    5  &   12  &   12  \\
	& \bi{0}{1}{1} &    2  &    2  &    3  &    3  \\
	& \bi{1}{0}{1} &    2  &    2  &    3  &    3  \\
	& \bi{1}{1}{0} &    2  &    2  &    3  &    3  \\
	& \bi{1}{1}{1} &    1  &    1  &    1  &    1  \\
	\cmidrule(r){1-2}\cmidrule(lr){3-4}\cmidrule(l){5-6}

	\multirow{8}{*}{3}
	& \bi{0}{0}{0} &  4,177  &  131  &  22,478  &  393  \\
	& \bi{0}{0}{1} &    164  &    0  &     503  &    0  \\
	& \bi{0}{1}{0} &    199  &    0  &     570  &    0  \\
	& \bi{1}{0}{0} &    145  &    0  &     455  &    0  \\
	& \bi{0}{1}{1} &     17  &    0  &      19  &    0  \\
	& \bi{1}{0}{1} &     18  &    0  &      18  &    0  \\
	& \bi{1}{1}{0} &     16  &    0  &      26  &    0  \\
	& \bi{1}{1}{1} &      0  &    0  &       0  &    0  \\
	\cmidrule(r){1-2}\cmidrule(lr){3-4}\cmidrule(l){5-6}

	\multirow{8}{*}{4}
	& \bi{0}{0}{0} &  102,086  &  2,536  &  9,700,876  &  17,867  \\*
	& \bi{0}{0}{1} &      596  &    596  &      1,422  &   1,422  \\*
	& \bi{0}{1}{0} &      515  &    515  &      1,348  &   1,348  \\*
	& \bi{1}{0}{0} &      414  &    414  &      1,106  &   1,106  \\*
	& \bi{0}{1}{1} &       49  &     49  &         95  &      95  \\*
	& \bi{1}{0}{1} &       52  &     52  &         99  &      99  \\*
	& \bi{1}{1}{0} &       48  &     48  &        116  &     116  \\*
	& \bi{1}{1}{1} &       14  &     14  &         22  &      22  \\

	\multirow{8}{*}{5}
	& \bi{0}{0}{0} &  1,490,304  &  19,708  &  16,489,333,269  & 271,705  \\*
	& \bi{0}{0}{1} &     29,020  &       0  &       5,949,935  &       0  \\*
	& \bi{0}{1}{0} &     35,504  &       0  &       7,060,847  &       0  \\*
	& \bi{1}{0}{0} &     23,500  &       0  &       4,338,210  &       0  \\*
	& \bi{0}{1}{1} &      2,137  &       0  &          16,976  &       0  \\*
	& \bi{1}{0}{1} &      1,630  &       0  &          11,711  &       0  \\*
	& \bi{1}{1}{0} &      1,488  &       0  &          10,386  &       0  \\*
	& \bi{1}{1}{1} &          0  &       0  &               0  &       0  \\*
	\cmidrule(r){1-2}\cmidrule(lr){3-4}\cmidrule(l){5-6}

	\multirow{8}{*}{6}
	& \bi{0}{0}{0} &  15,528,617  &  181,042  &  76,291,672,688,418  &  16,885,497  \\
	& \bi{0}{0}{1} &      25,193  &   25,193  &             209,150  &     209,150  \\
	& \bi{0}{1}{0} &      24,225  &   24,225  &             210,513  &     210,513  \\
	& \bi{1}{0}{0} &      19,267  &   19,267  &             163,841  &     163,841  \\
	& \bi{0}{1}{1} &       1,308  &    1,308  &               4,330  &       4,330  \\
	& \bi{1}{0}{1} &       1,342  &    1,342  &               4,263  &       4,263  \\
	& \bi{1}{1}{0} &       1,194  &    1,194  &               4,363  &       4,363  \\
	& \bi{1}{1}{1} &         103  &      103  &                 184  &         184  \\
	\cmidrule(r){1-2}\cmidrule(lr){3-4}\cmidrule(l){5-6}

	\multirow{8}{*}{7}
	& \bi{0}{0}{0} &  126,001,201  &  889,914  &  837,761,896,860,221,208  &  738,006,485  \\
	& \bi{0}{0}{1} &    1,515,854  &        0  &        1,470,062,260,025  &            0  \\
	& \bi{0}{1}{0} &    1,904,741  &        0  &        1,844,557,789,203  &            0  \\
	& \bi{1}{0}{0} &    1,300,827  &        0  &        1,038,215,861,595  &            0  \\
	& \bi{0}{1}{1} &       74,372  &        0  &               96,867,288  &            0  \\
	& \bi{1}{0}{1} &       56,077  &        0  &               67,359,392  &            0  \\
	& \bi{1}{1}{0} &       57,255  &        0  &               60,794,503  &            0  \\
	& \bi{1}{1}{1} &            0  &        0  &                        0  &            0  \\
	\cmidrule(r){1-2}\cmidrule(lr){3-4}\cmidrule(l){5-6}

	\multirow{8}{*}{8}
    & \bi{0}{0}{0} &  890,565,159  &  5,554,303  &  20,114,075,740,796,013,040,830  &  134,474,650,261  \\
    & \bi{0}{0}{1} &      478,934  &    478,934  &                      95,430,361  &       95,430,361  \\
    & \bi{0}{1}{0} &      453,539  &    453,539  &                      72,703,621  &       72,703,621  \\
    & \bi{1}{0}{0} &      371,308  &    371,308  &                      62,928,907  &       62,928,907  \\
    & \bi{0}{1}{1} &       19,057  &     19,057  &                         227,872  &          227,872  \\
    & \bi{1}{0}{1} &       16,456  &     16,456  &                         250,768  &          250,768  \\
    & \bi{1}{1}{0} &       14,148  &     14,148  &                         202,210  &          202,210  \\
    & \bi{1}{1}{1} &        1,658  &      1,658  &                           5,899  &            5,899
\end{longtable}


\newpage
\bibliography{braneCount}

\end{document}